\documentclass[sn-mathphys,Numbered]{sn-jnl}

\usepackage{graphicx}%
\usepackage{multirow}%
\usepackage{amsmath,amssymb,amsfonts}%
\usepackage{amsthm}%
\usepackage{mathrsfs}%
\usepackage[title]{appendix}%
\usepackage{xcolor}%
\usepackage{textcomp}%
\usepackage{manyfoot}%
\usepackage{booktabs}%
\usepackage{algorithm}%
\usepackage{algorithmicx}%
\usepackage{algpseudocode}%
\usepackage{listings}%

\begin{document}
\title[Laplacian Coarse Graining in Complex Networks]{Laplacian Coarse Graining in Complex Networks}

\author[1]{Matheus de C. Loures}\email{mloures@ifi.unicamp.br}
\author[1]{Alan Albert Piovesana}
 \email{alan.albert.p@gmail.com}
\author*[1]{José Antônio Brum}
 \email{ brum@unicamp.br}
\affil[1]{
Department of Condensed Matter Physics, Institute of Physics ``Gleb Wataghin'',  University of Campinas, Campinas - SP, Brazil.
}

\date{\today}

\abstract{

 Complex networks can model a range of different systems, from the human brain to social connections. Some of those networks have a large number of nodes and links, making it impractical to analyze them directly. One strategy to simplify these systems is by creating miniaturized versions of the networks that keep their main properties. A convenient tool that applies that strategy is the renormalization group (RG), a methodology used in statistical physics to change the scales of physical systems. This method consists of two steps: a coarse grain, where one reduces the size of the system, and a rescaling of the interactions to compensate for the information loss. This work applies RG to complex networks by introducing a coarse-graining method based on the Laplacian matrix.
 We use a field-theoretical approach to calculate the correlation function and coarse-grain the most correlated nodes into super-nodes, applying our method to several artificial and real-world networks. The results are promising, with most of the networks under analysis showing self-similar properties across different scales.
}
\keywords{Complex networks, renormalization, network laplacian, scaling}
\maketitle

\section{\label{sec:level1}Introduction}

The development of experimental techniques and computational methods  has allowed the study of systems with large numbers of interacting components. Network theory has developed in the last decades as one of the most popular tools to deal with such systems, that range from social interaction studies, such as scientific collaborations, communication networks, such as the internet, and biological systems - neuron networks of the brain - among several other examples (for a review, see \cite{newman_book, barabasi_book, bullmore_book}).

Many of these systems have a large number of interacting components, which brings a few problems: one is the experimental resolution in some scales of interaction since, for example, probing all neuron connections in brain networks is not currently feasible. Another problem is computational complexity since several algorithms that run on networks scale fast with the number of nodes and links. Yet another problem is that a large number of components can make it more difficult to interpret data and draw insight directly from the original network. To solve these issues, one can take inspiration from statistical physics to create methods to understand the structure and dynamics of  large   systems.

One strategy to tackle this problem is to reduce the network size to have a smaller, more tractable system.  The goal is, therefore, to obtain reduced versions of the network that preserve the main features of the original system - scale-invariant properties. 

Statistical physics offers a powerful method to deal with different scales of physical systems - the Renormalization Group (RG) theory \cite{wilson_1974, wilson_1983, kadanoff}. Developed in the context of critical phenomena, RG addresses a variety of physical systems to reduce the number of parameters used to describe them - creating miniaturized versions that preserve the important properties of the original problems. Consequently, the adaptation of this theory to complex networks has been the subject of intense research in the last years \cite{newman_watts, song_2005, song_2006, gfeller_2007, rozenfeld, erzan_2011, erzan_2015, bialek_2016, bialek_2018_2, bialek_2018, garcia-perez, garcia-perez-2, bianconi_2020, lahoche_2021, lahoche_2022, boguna_2021, garuccio, villegas_2022a, villegas_2022b}. This is, however, not straightforward because complex networks lack a proper geometric length scale to serve as a basis for coarse-graining the system. Even for spatial networks embedded in metric spaces, this is a significant difficulty since the interplay between the spatial embedding of the network and its topological properties is not always clear \cite{barthelemy}.  

The first known attempts to address the problem of coarse-graining a network used $k$-core decomposition \cite{bollobas}, a decimation approach to reduce the number of nodes, or, community detection methods \cite{girvan_newman} to reduce the size of the network by clustering the nodes into super-nodes. Song et al. \cite{song_2005, song_2006} proposed a coarse-graining scheme for networks based on fractal dimension, following a box-counting method. On another path, García-Pérez et al. \cite{garcia-perez} developed an RG approach based on geometric representations. In their method, the network is embedded into an underlying metric space, creating hidden variables that allow for direct coarse-graining of blocks into super-nodes. Another approach that uses metric spaces applies internal dynamic processes on the network to coarse-grain it, and it is associated with spectral properties of the network \cite{gfeller_2007}.   Recently, Villegas et al. \cite{villegas_2022a, villegas_2022b} proposed a Laplacian RG diffusion method to coarse-grain the network into super-nodes. All these methods closely follow the block-spin renormalization method developed by Kadanoff \cite{kadanoff_2000}.  Aygün and Erzan \cite{erzan_2011} and Tuncer and Erzan \cite{erzan_2015}  proposed a different approach in which the authors developed a spectral renormalization for networks inspired by the momentum space RG developed for statistical physics by Wilson \cite{wilson_1974, wilson_1983}. Furthermore, Bialek and collaborators \cite{bialek_2016, bialek_2018_2, bialek_2018} suggested a strong analogy between RG theory and principal component analysis (PCA), a method used to analyze high-dimensional data. Diagonalization of the covariance matrix allows one to find the most relevant variables and exclude the others from the analysis. In some cases, a few variables are sufficient to describe most of the properties in the data. In particular, the authors applied their model \cite{bialek_2018_2} to a  time-dependent neuron activity network. Recently, Lahoche et al. \cite{lahoche_2021, lahoche_2022} further developed these ideas proposing a field-theoretical RG approach for data analysis.

\begin{center}\
\begin{figure}[H]\
    \includegraphics[width=1.0\textwidth]{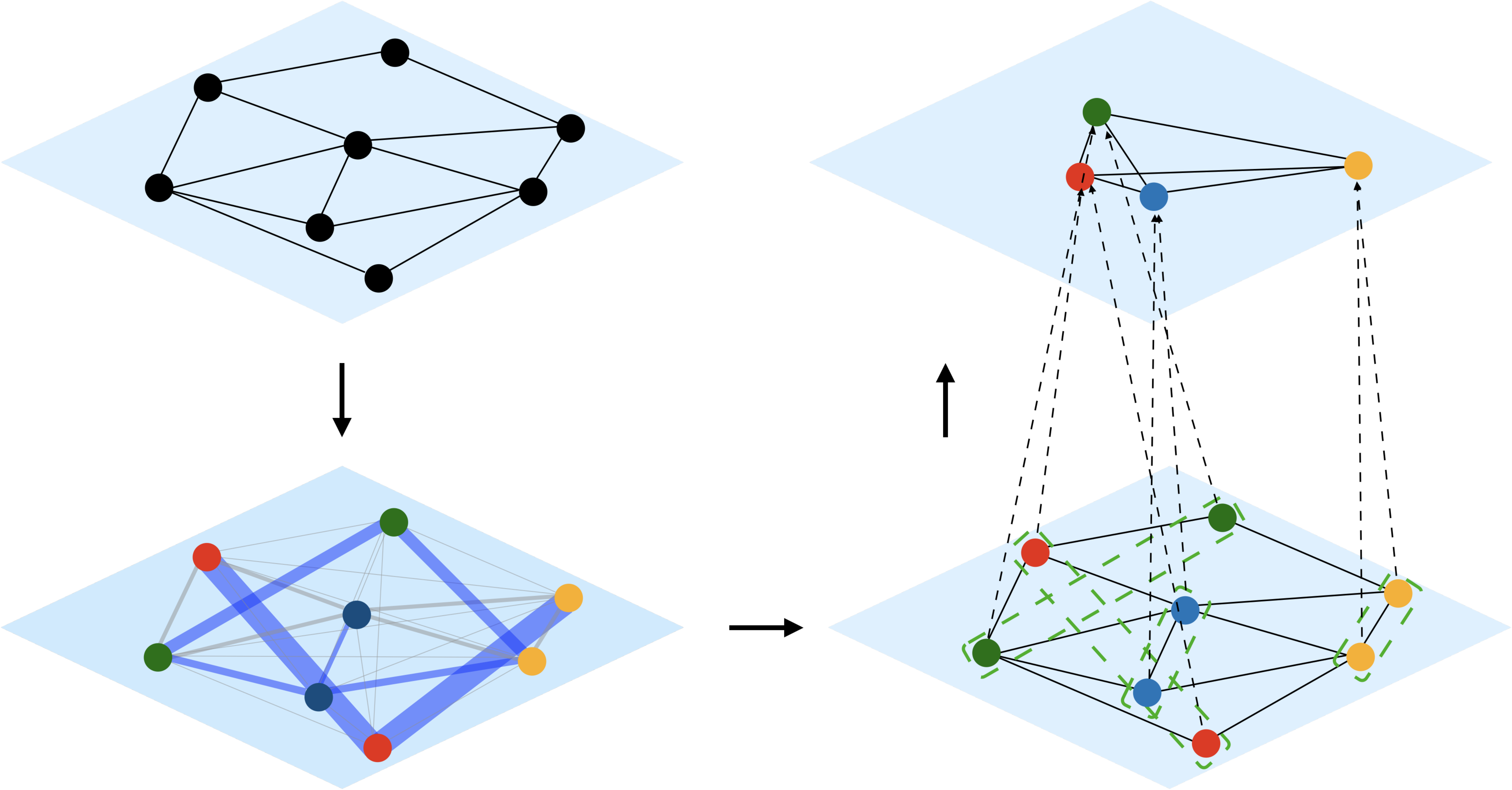}\hfill

    \caption{(Color online.) Scheme for the Laplacian coarse graining.  Counterclockwise, we have an arbitrary network with eight nodes and thirteen links (left upper panel), followed by the correlation   between  the nodes (left lower panel) with     positive correlations in blue and lines thickness proportional to the correlation value and negative correlations in gray with  lines thickness proportional to the inverse of the absolute value, with nodes colored according to the super nodes to be formed, followed by the network showing the formation of the super nodes (right lower panel) and finally the renormalized network (right upper panel).}

    \label{fig:renorm}
\end{figure}\
\end{center}\
In this work, we propose an alternative method to coarse-grain the network extending the \textit{direct correlations} approach of Meshulam et al.\cite{bialek_2018_2} to structural networks and inspired by a field-theoretical description of the network dynamics \cite{lahoche_2022,bianconi_2020, villegas_2022b}. The authors proposed to coarse-grain the most correlated nodes pairwise into super-nodes. Correlation between nodes was obtained from time-dependent neuron activity. Our goal is to obtain a coarse-graining scheme for  structural networks. We do this by associating the correlation matrix with the network Laplacian to look into the relationship between the structure of the network and its dynamic processes. Figure \ref{fig:renorm} shows our coarse grain methodology schematically.

The paper is structured as follows. In section \ref{chap2}, we describe our method for the coarse grain of complex networks. In section \ref{chap3}, we apply it to artificial and real-world networks. Finally, we make our final remarks in section \ref{chap4}.

\section{Laplacian approach for coarse-graining in complex networks}
\label{chap2}

We search for a method to coarse-grain the network based on its structural properties by adapting the direct correlations coarse-graining model developed by L. Meshulam et al. \cite{bialek_2018_2}. 

The basic idea is to form super-nodes by coarse-graining nodes that are more strongly correlated. This procedure loosely followed the model proposed by Fisher \cite{fisher_1995} on strongly disordered systems.  Meshulam et al. \cite{bialek_2018_2} applied their model to a neuron network where the associated variable was the time-dependent neuron activity, and  they  obtained the covariance matrix from the correlation between those activities. Neurons with the most strongly correlated activities were then grouped together. They identified the largest off-diagonal element of the correlation matrix and coarse-grained the two neurons into super-nodes. After that, they looked at the second largest correlation that did not involve the same $i$ and $j$ of the first step, coarse-grained, and repeated, summing up the activities of the nodes that formed the super-nodes to obtain a new covariance matrix.

We are interested in coarse-graining the structural network, which suggests using the field-theoretical correlation function, $\left<\phi_i\phi_j\right>-\left<\phi_i\right>\left<\phi_j\right>$, in which $\left<.\right>$ denotes the ensemble average, using the discrete Laplacian matrix to obtain a coarse-graining procedure for the structural network. To compute the correlation function,  we follow a similar framework as the one proposed by Meshulam and co-authors  \cite{bialek_2018_2} and Lahoche et al. \cite{lahoche_2021, lahoche_2022}.

An undirected binary network represents the complex system under analysis. It is expressed by the adjacency matrix $A_{ij}$, whose elements are equal to one when there is a link between nodes $i$ and $j$, and zero otherwise. We take the network nodes as our variables and associate to the node $i$ a state represented by $\phi_i$. The full state of the network is represented by $|\Phi\rangle=|\phi_1, \phi_2, ...,\phi_N\rangle$,   and it  is then described by the probability distribution

\begin{equation}
    p[\Phi]=\frac{1}{Z}e^{-\mathcal{H}[\Phi]},
\end{equation}
where $\mathcal{H}[\Phi]$ represents the functional Hamiltonian, and $Z$ is the normalization term, equal to the partition function:

\begin{equation}
      Z(J)=\int d\Phi \, e^{-\mathcal{H}(J)}.
      \label{eq:partition}
\end{equation}

We integrate over all possible states of the network, that is, $d\Phi=d\phi_1 d\phi_2 ...d\phi_N$.   

In the framework of  Landau's model \cite{huang,burioni_2005,burioni_1999,watanabe_1987,burioni_2005_2}, the functional Hamiltonian reads: 

\begin{equation}
    \mathcal{H}= \int d\boldsymbol{x}\Big[\frac{1}{2}\Big[|\nabla \phi(\boldsymbol{x})|^2 +r_0\phi^2( \boldsymbol{x})\Big]-J\phi(\boldsymbol{x})\Big],
    \label{hamiltonian1}
\end{equation}where the first term represents the energy of the system, and the last one includes an external field acting on it. We limit ourselves to second-order approximation for the Gaussian model and do not include higher-order interactions among the nodes.

Gaussian models were previously studied in networks \cite{burioni_2005,burioni_1999,watanabe_1987,burioni_2005_2,estrada_hatano_2008, estrada_hatano_2010, estrada_hatano_2012}. We connect this model to our network by associating the position $\boldsymbol{x}$ to a discrete variable representing node $i$. The value of the state $\phi(\boldsymbol{x})$ is the state at node $i$, that is, using the map $\phi(\boldsymbol{x})\rightarrow \phi_i$. The integral over the space coordinates now becomes a sum over the nodes. We do not have physical distances but instead, topological ones. Following references  \cite{erzan_2011} and \cite{burioni_2005_2,luxburg_2007}, we write

\begin{equation}
    \int d \boldsymbol{x} |\nabla \phi(\boldsymbol{x})|^2 \rightarrow \frac{1}{2}\sum_{ij} A_{ij} (\phi_i-\phi_j)^2
\end{equation}
where the factor $1/2$ is to compensate for the double counting.  The Hamiltonian \eqref{hamiltonian1} then reads:

\begin{equation}
   \mathcal{H}=\frac{1}{2}\sum_{ij} \left(\frac{1}{2}A_{ij}(\phi_i-\phi_j)^2 + r_0\phi_i^2\delta_{ij} \right)-\sum_i J_i\phi_i.
\end{equation}

 It is straightforward to obtain

\begin{align}
    \mathcal{H}=\frac{1}{2} \sum_{ij} \left(  \phi_i(k_i \delta_{ij}-A_{ij})\phi_j +r_0 \phi_i^2\delta_{ij} \right)-\sum_i J_i\phi_i,
\end{align} where $k_i=\sum_j A_{ij}$ is the degree of the $i^{th}$ node. We identify $L_{ij}=k_i \delta_{ij}-A_{ij}$ as the Laplacian matrix for the network \cite{newman_book}. Finally, we have $\mathcal{H}$ as:

\begin{equation}
     \mathcal{H}=\frac{1}{2}\sum_{ij}   \phi_i \left(L_{ij} + r_0\delta_{ij}\right)\phi_j-\sum_iJ_i\phi_i.
     \label{eq:H}
 \end{equation}

 With the partition function (equation \ref{eq:partition}), computed using equation \ref{eq:H}, we can calculate the correlation function. It is possible to show that (details in Appendix \ref{appendix:a}):
\begin{align}
        \langle \phi_i \rangle&=\frac{\partial}{\partial J_i} \ln(Z[J])\Big|_{J=0} =\frac{1}{Z}\int d\Phi \phi_i e^{-\mathcal{H}[J]}
        \label{eq:av_phi}
\end{align}

and
\begin{align}
     \frac{\partial^2\ln( Z)}{\partial J_i \partial J_j}&=  -\langle \phi_i \rangle \langle \phi_j \rangle +\frac{1}{Z} \frac{\partial^2Z}{\partial J_i \partial J_j} 
     \label{eq:av_phi_2}
\end{align}

And since
\begin{align}
    \frac{1}{Z} \frac{\partial^2Z}{\partial J_i \partial J_j}
    &= \frac{1}{Z}\int d\Phi (\phi_i \phi_je^{-\mathcal{H}[J]})= \langle \phi_i \phi_j\rangle,
    \label{eq:av_phi_3}
    \end{align}
we finally obtain the correlation function:

\begin{equation}
    \frac{\partial^2\ln( Z)}{\partial J_i \partial J_j}=\langle \phi_i \phi_j \rangle - \langle \phi_i \rangle  \langle \phi_j \rangle=c_{ij}.
    \label{eq:cov}
\end{equation}

We proceed using Gaussian integrals. For simplicity let us write $K_{ij}=\frac{1}{2}(L_{ij}+r_0\delta_{ij})$. $K$ is a symmetric matrix with positive eigenvalues ${\lambda_i}$ for undirected networks. Let us diagonalize $K$ with $\Lambda$ being the transformation matrix that makes $K$ diagonal and $D$ the diagonal matrix formed by its eigenvalues. $K$ is orthonormal, so its inverse is the transposed matrix. Let us write $\phi = (\phi_1, \phi_2, ...,\phi_N)^T$ and $J=(J_1,J_2,...,J_N)^T$, that is, $\phi$ and $J$ are two column vectors. The partition function is then

\begin{equation}
    Z[J]=\prod_k\int_{-\infty}^{\infty}d\phi_k e^{-\phi^T\Lambda^TD\lambda \phi+J^T\phi}
    \label{z_j_d}
\end{equation}

Developing the previous equation (see Appendix \ref{appendix:a} for details) we obtain

\begin{align}
    \frac{\partial^2 \ln(Z)}{\partial J_k \partial J_l}= \frac{1}{2} K^{-1}_{kl}
    \label{eq:z_k_corr}
\end{align}

Therefore, the correlation function is half the inverse matrix element that couples $\phi_i$ and $\phi_j$, and we have the correlation matrix being equal to:
\begin{equation}
    {c}=\frac{1}{2}K^{-1}=\frac{1}{L+r_0\mathbb{I}},
\end{equation}
where $\mathbb{I}$ is the identity matrix.

It is important to notice that $L$ is singular due to the null eigenvalue associated with the eigenvector $(1,1,1,...,1)^T$, from which we have two options to calculate the correlation matrix: choosing a non-null positive value for $r_0$ to make $c$ invertible or computing the pseudo-inverse of the Laplacian matrix \cite{rand_walk}. In the first case, since we are simply adding the identity times a constant to the Laplacian, we can choose any value for $r_0$, and the Laplacian matrix will determine the correlations between the nodes. Both methods - using $r_0$ or the pseudo-inverse - have the same ordering of eigenvalues  and eigenvectors \cite{rand_walk} except for the eigenvalue $0$. In this work, we use the pseudo-inverse of the Laplacian in our calculations, identifying it with the correlation matrix.

Once we obtain the correlation matrix, we coarse-grain the nodes ordered by the correlation values following Meshulam et al. \cite{bialek_2018_2} model. For that, we form a new node combining the two most correlated nodes and iterating. We call the coarse-grained node a “super-node" and refer to each level of coarse-graining by the index $l$. To build a rescaled network, we need to define the links between the super-nodes, and for a binary network, we ignore self-loops and add a link between two super-nodes if there are one or more links between nodes inside one super-node and the other.

The pseudo-inverse of the Laplacian has been the subject of study in the context of nodes communicability \cite{rand_walk,estrada_hatano_2008, estrada_hatano_2010, estrada_hatano_2012}. Fouss et al. \cite{rand_walk} have identified it with a similarity measure and related to the average commute time between nodes $i$ and $j$. Estrada and Hatano \cite{estrada_hatano_2008, estrada_hatano_2010, estrada_hatano_2012} have built a communicability function directly associated with the pseudo-inverse of the Laplacian. In their work, the pseudo-inverse of the Laplacian is associated with the thermal Green function of a network of harmonic oscillators where the nodes are balls with mass $m$ and links are springs.  Conceptually, the thermal Green function determines the correlations between nodes' displacement due to thermal fluctuations in the network of harmonic oscillators. The pseudo-inverse of the Laplacian in that model is interpreted as a measure of communicability because it tracks the response of the nodes to external perturbations and how the displacement of each node affects the others.

\section{Results}
\label{chap3}

We test our coarse-grain model by examining a few topological  properties of the network.  We first analyze artificial networks and then apply the method to five real networks \cite{github_1}.

\subsection{Coarse-graining  for  artificial networks}

 We begin by applying our model to artificial networks. That allows us to understand the effects of our coarse-graining within random networks with well-known behavior. We use an Erdos-Rényi network with a probability of connection $p=0.01$ and a Barabási-Albert network with preferential attachment parameter $m=5$, both with $N=1024$ nodes. It is important to observe that we are coarse-graining the network but preserving all possible links in a binary network. As a consequence, in general, the average degree increases. To compare the different coarse-grained network properties, we rescale the network degree, that is $k^{(l)}\rightarrow k^{(l)}/\langle k^{(l)}\rangle$, where $\left< k^{(l)} \right>$ is the average network degree at the $l^{th}$ level of rescaling \cite{garcia-perez}.

We analyze the coarse-graining results by computing the complementary cumulative degree distribution, degree-degree correlation, and degree-dependent clustering for rescaled degree classes.  These parameters have been used in the works on geometric renormalization  \cite{garcia-perez, garcia-perez-2} and allow  us to verify the similarity of nodes correlation and network structural organization as we downscale the network.  Figure \ref{fig:ER_BA_plot_all} shows the results for Erdos-Rényi (ER) (left panels)  and the Barabási-Albert (BA) (right panels) networks. The degree-degree correlation, $k_{nn,n}$, is computed by the normalized average degree for the nearest-neighbors i.e. $k_{nn,n}(k^{(l)}/ \langle k^{(l)} \rangle)=k_{nn}(k^{(l)}/\langle k^{(l)} \rangle) \frac{\langle k^{(l)} \rangle}{\langle (k^{(l)})^2 \rangle}$ \cite{garcia-perez}.  
We observe that the coarse-graining preserves most of the properties for both the Erdos-Rényi and Barabási-Albert networks. The complementary cumulative degree distribution collapses into a single curve.  The degree-degree correlation shows values close to unit almost regardless of the rescaled degree. This means that the network downscaling obtained by our coarse-graining method preserves the random behavior of the networks. For this reason, we show the degree-dependent clustering for ER and BA networks compared to the uncorrelated network clustering  value $c_0=(\langle (k^{(l)})^2\rangle-\langle k^{(l)}\rangle)^2/(N\langle k^{(l)}\rangle^3)$ \cite{boguna_2003, newman_2003}. We observe that all levels of coarse-grain collapse into a single curve, which is approximately constant and close to one. These results confirm the random nature of both BA and ER networks with no significant difference as we downscale the networks \cite{vespignani_2007}. 

The random nature of the networks manifests itself in the pseudo-inverse of the Laplacian (see Appendix \ref{appendix:b}, fig. \ref{fig:pinv_art}). We observe that its random structure is preserved as we downsize the network, and it approximately follows the adjacency matrix behavior (not shown here). In particular, for the Barabási-Albert, the power-law degree distribution $\rho(k)=k^{-\gamma}$ is preserved, with $\gamma$ varying as table \ref{tab:table_gamma}. 

\begin{table}[ht!]
\caption{$\gamma$ values for Barabási-Albert network.}
\label{tab:table_gamma}
\begin{tabular}{ccccc}

Layers ($l$) & $  \gamma$ \\ \hline
\; \, $l=0$ & $2.72 \pm 0.01$\\ \hline
\; \,
$l=1$ & $3.03 \pm 0.06$ \\ \hline
\; \,
$l=2$ & $3.15 \pm 0.06$ \\ \hline
\; \,
$l=3$ & $3.36 \pm 0.08$ \\ \hline

\end{tabular}
\end{table}

\begin{center}\
\begin{figure}[H]\
    \centering
    \includegraphics[width=1.0\textwidth]{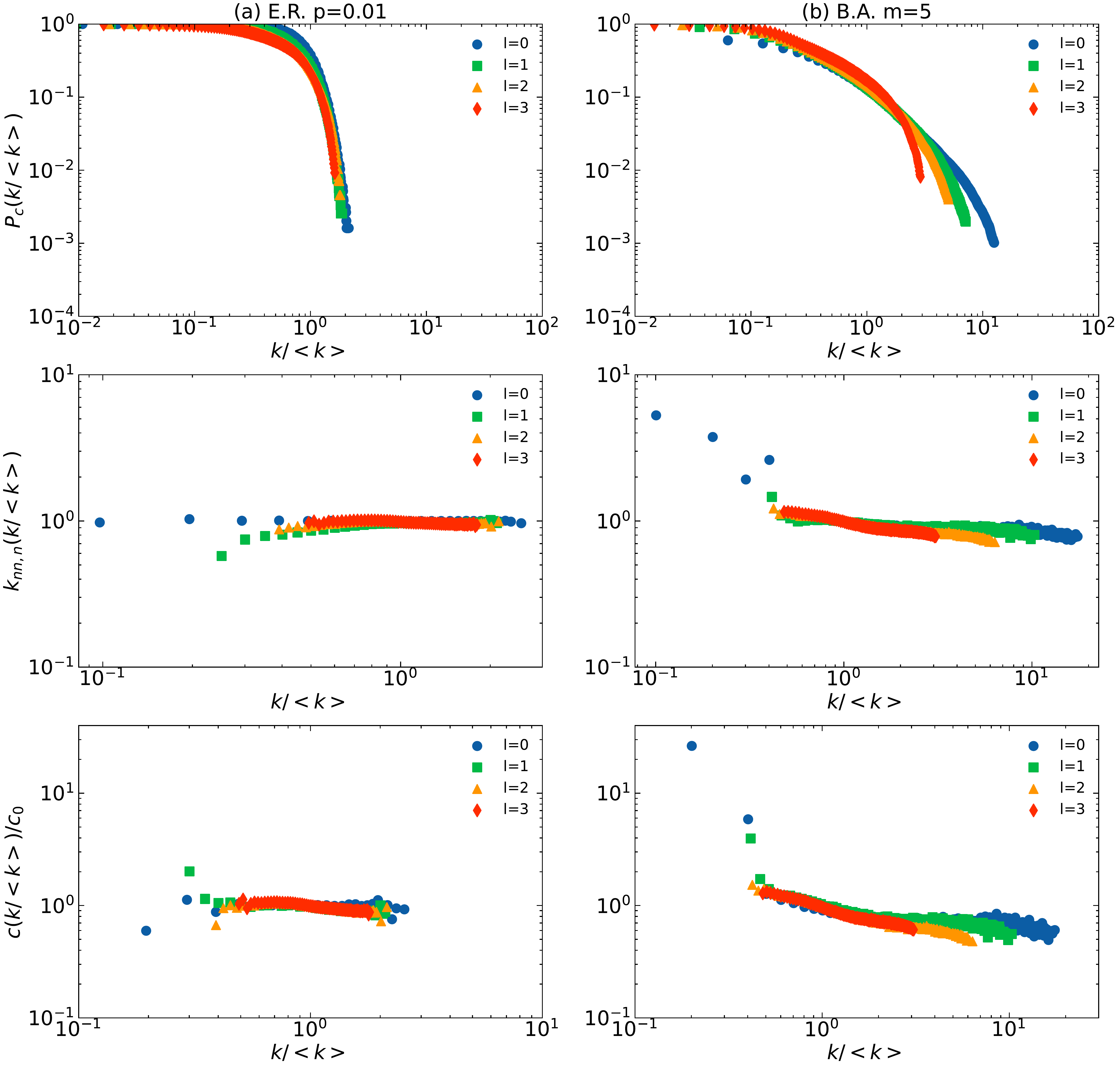}\hfill
    \caption{(Color online.) Comparison between the original network and three levels of coarse-graining (indicated by the upper index $(l)$) for Erdos-Rényi networks with $N=1024$ and $p=0.01$, and Barabási-Albert networks with  $m=5$ and the same number of nodes.  The topological properties shown are: complementary cumulative distribution, degree-degree correlation and the degree-dependent clustering compared to its value for uncorrelated networks. The above results were averaged over 100 networks of each type.} 
    \label{fig:ER_BA_plot_all}
\end{figure}\
\end{center}\

\subsection{Coarse-graining for  real networks}

We now examine the results of Laplacian coarse graining in real networks. We study the networks of the connectome of the nematode \textit{C. elegans} \cite{wormatlas}, the US Airport traffic (Airport) \cite{airport,colizza,opsahl},  the  HI-II-14 human interactome (Proteome) \cite{proteome}, the words adjacency network from Darwin's book ``The Origin of Species'' (Words) \cite{words}, and the US Power Grid network (PowerGrid) \cite{powergrid}.  

Table \ref{tab:table_real_1} contains the parameters that characterize the networks.  In Fig. \ref{fig:real_plot_all1} and Fig. \ref{fig:real_plot_all2}, we show the results for the original network and the three first levels of coarse-graining for the complementary cumulative degree distribution, degree-degree correlation, and degree-dependent clustering for rescaled degree classes.

\begin{table}
\caption{Properties of the real networks, where $N$ is the number of nodes, $\left< k \right>$ the average degree, $\left< c \right> $ the average node clustering, and $\left< L \right>$ is the mean path length.}
\label{tab:table_real_1}
\begin{tabular}{ccccc}
 
 Networks        & $\; \; N$        &  $\; \;  \langle k \rangle$  & $\; \;  \langle c \rangle$      &  $\; \;   \langle L \rangle$   \\ \hline
 
C.elegans        &  278       &         16.42     &       0.34            & 2.44      \\ \hline
Airport          &   2905     &         10.77     &       0.46            & 4.10        \\ \hline
US powergrid     &    4941    &         2.67      &       0.080            & 18.99          \\ \hline
Words            &    7377    &         11.98     &       0.41            & 2.78       \\ \hline
Proteome         &  3985      &           6.77   &       0.045           & 4.09         \\ \hline
\end{tabular}
\end{table}

\begin{center}\
\begin{figure}[H]\
    \includegraphics[width=1.0\textwidth]{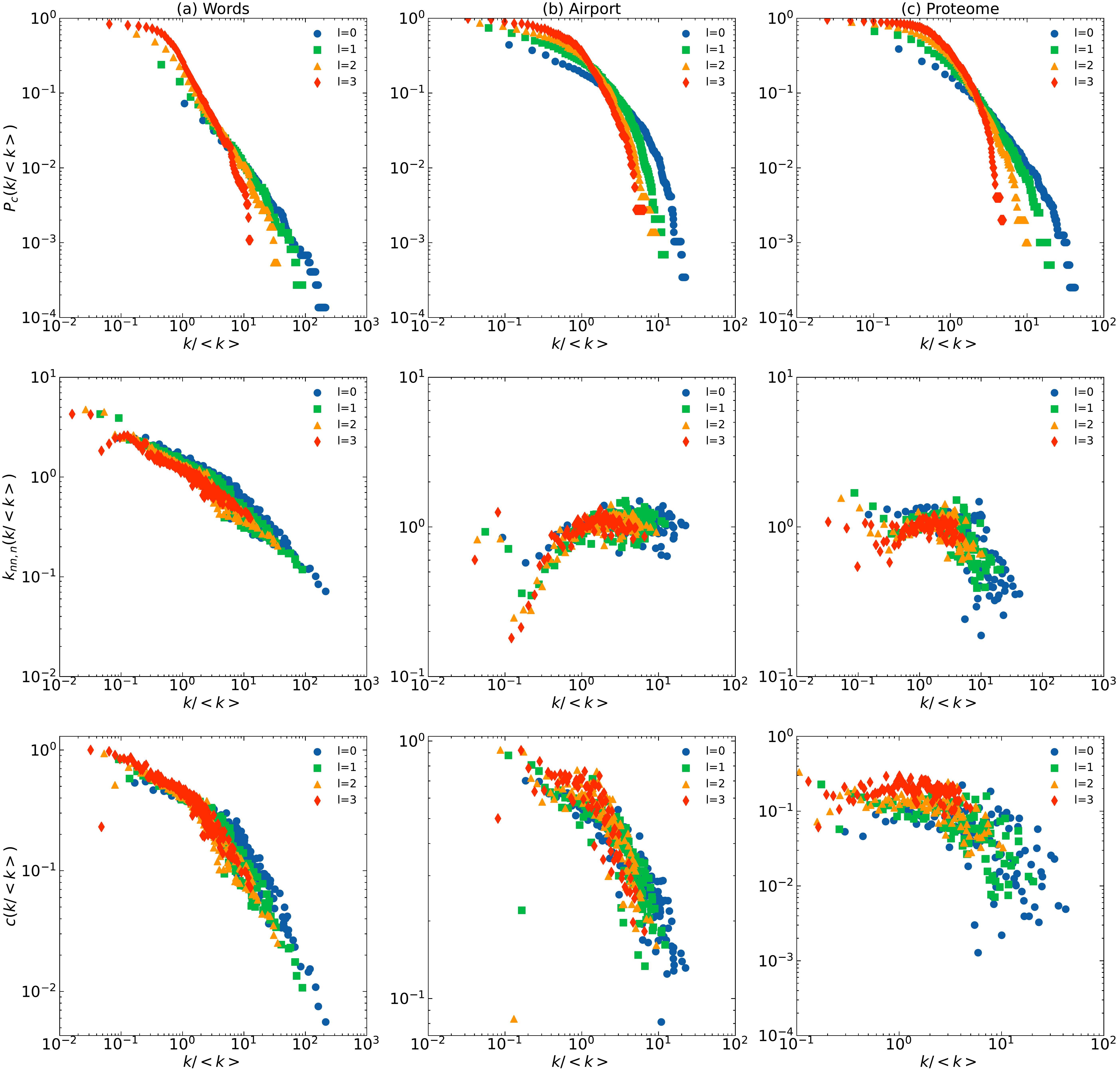}\hfill

    \caption{(Color online.) Comparison between the original network and three levels of coarse-graining (indicated by the upper index $(l)$) for the networks of Words, Airport, and Proteome for the topological properties complementary cumulative distribution, degree-degree correlation and degree-dependent clustering.}
    \label{fig:real_plot_all1}
\end{figure}\
\end{center}\

\begin{figure}[H]\
\centering
    \includegraphics[width=1.0\textwidth]{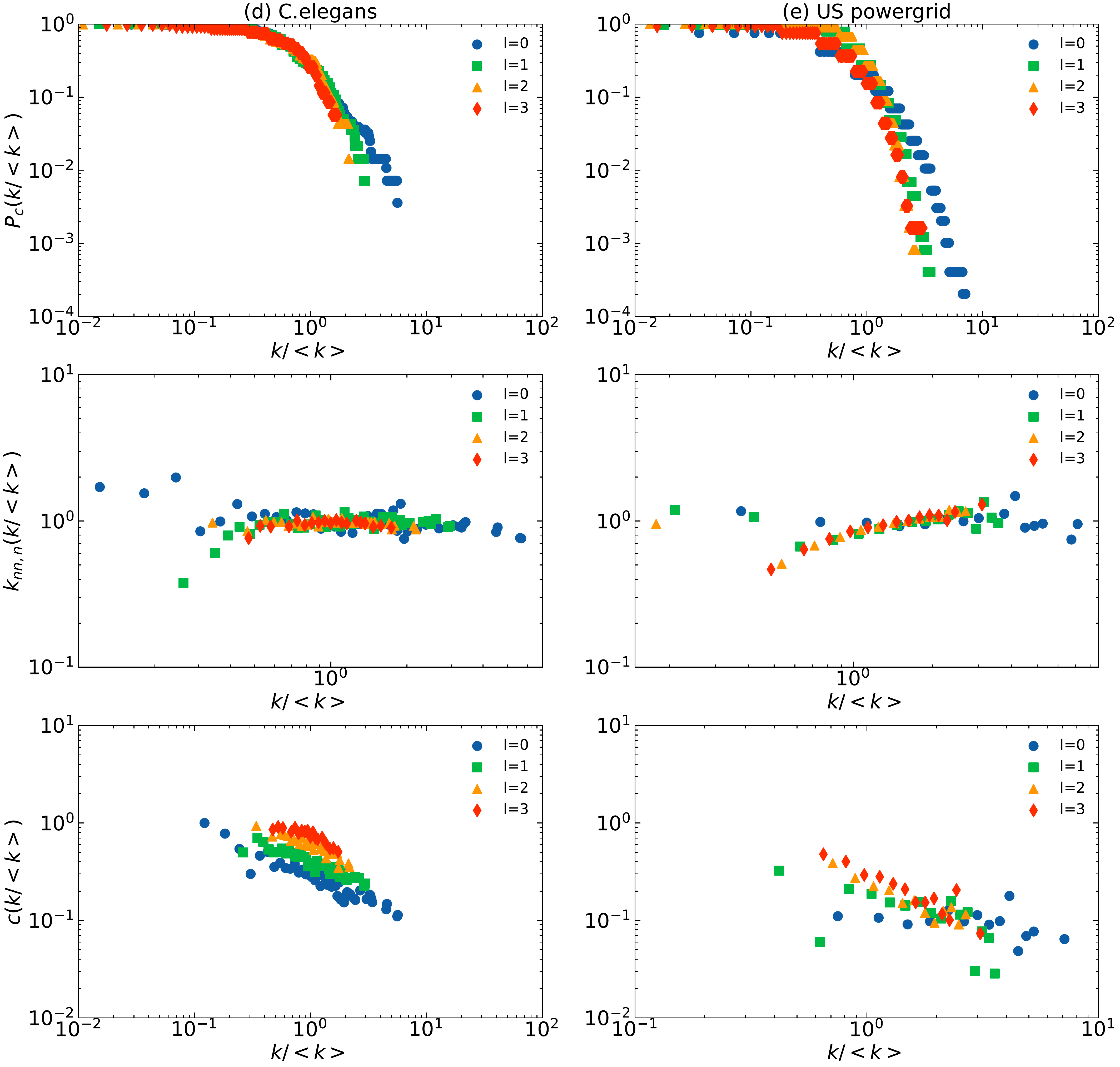}\hfill

    \caption{(Color online.) Comparison between the original network and three levels or coarse-graining (indicated by the upper index $(l)$) for the networks of the nematode \textit{C.elegans} and PowerGrid for the same properties as in Fig. \ref{fig:real_plot_all1}. }
    \label{fig:real_plot_all2}
\end{figure}\

We observe self-similarity for the complementary cumulative degree distribution with more consistent results for \textit{C. elegans}, and Words networks. The degree-degree correlation shows self-similarity for \textit{C. elegans}, Words, Proteome, and  Airport networks, with the curves collapsing into a single one, except for lower values of $k$ where the data show strong dispersion. For the Power Grid network, we observe a change in behavior as we downscale the network with degree-degree correlation increasing with  the rescaled degree. \textit{C. elegans} network shows an almost uncorrelated result for the degree-degree correlation. For the Airport, Proteome, and Words networks, we observe similar behavior for the degree-dependent clustering across all coarse-grained networks.  \textit{C. elegans} network shows hints of  self-similar behavior but with a slight shift to higher values, as we downscale the network. This reflects the weak degree-degree correlation of the network.  The degree-dependent clustering behavior for the Power Grid network does not show self-similarity, instead, it shows an increasing negative correlation with the rescaled degree as we downscale the network. The topological structure of the real networks can be observed in the pseudo-inverse of the Laplacian matrix (see Appendix \ref{appendix:b}, fig. \ref{fig:pinv_real1} and fig. \ref{fig:pinv_real2}), which shows a behavior that is similar to the adjacency matrix (not shown here). The coarse-grain rescaling obtained by the network preserves these structures up to a certain level of downscaling.

These results suggest that our method for network coarse-graining preserves the network characteristics for a certain class of networks. Airport, Words, \textit{C. elegans} and Proteome networks show a significant level of self-similarity as we coarse grain the networks. \textit{C. elegans} has a slight divergence in the degree-dependent clustering associated to its weak degree-degree correlation. The results, however, are less satisfactory for the Power Grid.

Comparing these results with the ones obtained by geometric renormalization \cite{garcia-perez}, the behaviors of the Airport, Words, Proteome, and \textit{C. elegans} networks are similar to some of their results.  The properties of the Power Grid network that lead to divergent results are not fully clear and are subject of investigation.      

\section{Conclusion}
\label{chap4}

In this work, we have introduced a methodology for coarse-graining networks to obtain rescaled versions of them. Our method is inspired by Bialek and collaborators \cite{bialek_2018_2, bialek_2018}, and it uses the network Laplacian to obtain a correlation matrix, rescaling the nodes into super-nodes. We applied it to artificial and real-world networks. The results are promising, preserving the original structures for all the artificial networks and most of real-world ones. Further work is necessary to identify the class of systems in which our method performs best.

To fully renormalize the networks, the algorithm also needs to rescale the interactions of the systems. That means, for undirected binary networks, finding a methodology for pruning the links. Garcia-Pérez \cite{garcia-perez} suggested a pruning method to obtain a network replica preserving the average degree, and it successfully replicates the dynamical behavior of the original networks. On another path, Villegas et al. \cite{villegas_2022b} developed a complete renormalization procedure that also rescales the links. Future work is necessary to develop a proper way to fully renormalize the network with our method preserving its binary feature.

\bmhead{Acknowledgments}
M. de Carvalho Loures acknowledges financial support from  CAPES through project number 88887.517253/2020-00 and CNPq through project number 403625/2022-0. A.A. Piovesana acknowledges financial support from SAE-PIBIC. We also thank A. Saa, and M.A.M. de Aguiar for their valuable additions to the discussions.

\clearpage

\begin{appendices}
\section{Detailed calculation of equations \ref{eq:cov} and \ref{eq:z_k_corr}: }
\label{appendix:a}

In this Appendix we prove some of the results used in the main text.

We start with the expression for the average of node state, $\left<\phi_i\right>$, eq. \ref{eq:av_phi}:

\begin{equation}
    \langle \phi_i \rangle= \frac{\partial}{\partial J_i}\ln(Z[J])
\end{equation}

Using $K_{ij}=\frac{1}{2}(L_{ij}+r_0\delta_{ij})$ for simplicity, the demonstration follows as

\begin{align}
        \langle \phi_i \rangle&=\frac{\partial}{\partial J_i} \ln(Z[J])\Big|_{J=0} =\frac{1}{Z}\int d\Phi \frac{\partial}{\partial J_i}e^{-\mathcal{H}[J]}\\&=\frac{1}{Z}\int d\Phi \frac{\partial}{\partial J_i}\exp\left(-\sum_{ij}\phi_i K_{ij}\phi_j+\sum_i J_i\phi_i\right)\\&=\frac{1}{Z}\int d\Phi \phi_i e^{-\mathcal{H}[J]}=\langle\phi_i\rangle.
\end{align}

We follow by obtaining the correlation function $\left<\phi_i\phi_j\right>-\left<\phi_i\right>\left<\phi_j\right>$:

\begin{equation}
    \frac{\partial^2\ln( Z)}{\partial J_i \partial J_j}=\langle \phi_i \phi_j \rangle - \langle \phi_i \rangle  \langle \phi_j\rangle.
\end{equation}

\begin{align}
     \frac{\partial^2\ln( Z)}{\partial J_i \partial J_j}&= \frac{\partial}{\partial J_i}\left( \frac{1}{Z} \frac{\partial}{\partial J_j}{Z}\right)\\ &=- \frac{1}{Z^2}\frac{\partial}{\partial J_i}{Z} \frac{\partial}{\partial J_j}Z +\frac{1}{Z} \frac{\partial^2Z}{\partial J_i \partial J_j}\\
     &=- \left(\frac{1}{Z}\frac{\partial}{\partial J_i}{Z}\right) \left( \frac{1}{Z}\frac{\partial}{\partial J_j}{Z}\right) +\frac{1}{Z} \frac{\partial^2Z}{\partial J_i \partial J_j} \\
     &=         - \Big(\frac{\partial}{\partial J_i}\ln Z\Big) \Big(\frac{\partial}{\partial J_j}\ln Z\Big) +\frac{1}{Z} \frac{\partial^2Z}{\partial J_i \partial J_j}    \\
     &= -\langle \phi_i \rangle \langle \phi_j \rangle +\frac{1}{Z} \frac{\partial^2Z}{\partial J_i \partial J_j} 
\end{align}

But,
\begin{align}
    \frac{1}{Z} \frac{\partial^2Z}{\partial J_i \partial J_j} &= \frac{1}{Z} \frac{\partial^2}{\partial J_i \partial J_j} \int d\Phi e^{-\mathcal{H}[J]}\\
    &= \frac{1}{Z} \frac{\partial^2}{\partial J_i \partial J_j} \int d\Phi e^{(-\sum_{i,j} \phi_i K_{ij} \phi_j + \sum_i J_i \phi_i )}\\
    &=\frac{1}{Z} \int d\Phi (\phi_i \phi_je^{-\sum_{i,j} \phi_i K_{ij} \phi_j + \sum_i J_i \phi_i })\\
    &=\frac{1}{Z} \int d\Phi (\phi_i \phi_je^{-H }) = \int d\Phi (\phi_i \phi_j\rho)= \langle \phi_i \phi_j\rangle
    \end{align}
    
Finally, we have

\begin{equation}
    \frac{\partial^2\ln( Z)}{\partial J_i \partial J_j}=\langle \phi_i \phi_j \rangle - \langle \phi_i \rangle  \langle \phi_j \rangle.
\end{equation}

We want to obtain an expression for the correlation function in terms of the network Laplacian (eq. \ref{eq:z_k_corr}). We start with the partition function, eq. \ref{z_j_d}: 

\begin{equation}
    Z[J]=\Pi_k\int_{-\infty}^{\infty}d\phi_k e^{-\phi^T\Lambda^TD\lambda \phi+J^T\phi}
\end{equation}

Let us write $\Lambda \phi= \psi $
\begin{align*}
    Z&= \; \prod_k\int_{-\infty}^{\infty}  d\psi_k det(\Lambda) e^{- \psi^T D \psi +  J^T \Lambda^T \psi } = \;  det(\Lambda)\prod_k\int_{-\infty}^{\infty}  d\psi_k  e^{- \psi^T D \psi +   (\Lambda J)^T \psi } \\
    &= \; det(\Lambda) \prod_k\int_{-\infty}^{\infty}  d\psi_k  e^{- \psi^T D \psi +   (\Lambda J)^T \psi } 
\end{align*}
defining $\Lambda J=h$,
\begin{align*}
    Z&= \;det(\Lambda) \prod_k\int_{-\infty}^{\infty}  d\psi_k  e^{- \psi^T D \psi +   h^T \psi } = \; det(\Lambda)\prod_k\int_{-\infty}^{\infty}  d\psi_k e^{- \sum_i \lambda_i \psi_i^2 +  \sum_i h_i \psi_i }\\ &= \; det(\Lambda)\prod_k\int_{-\infty}^{\infty}  d\psi_k e^{-  \lambda_k \psi_k^2 +   h_k \psi_k } = \; det(\Lambda)\prod_k \sqrt{\frac{\pi}{\lambda_k}} e^{\frac{h_k^2}{4\lambda_k}}\\
    &= \; det(\Lambda) \sqrt{\frac{(\pi)^N}{ \prod_k\lambda_k}} e^{ \sum_k\frac{h_k^2}{4\lambda_k}} \quad \quad \quad \; \; \, = \; det(\Lambda) \sqrt{\frac{(\pi)^N}{ det(K)}} e^\frac{h^T D^{-1}h}{4}\\
    &= \;
    det(\Lambda) \sqrt{\frac{(\pi)^N}{ det(K)}} e^\frac{(\Lambda J)^T D^{-1}\Lambda J}{4} \quad \; \; = \; det(\Lambda) \sqrt{\frac{(\pi)^N}{ det(K)}} e^\frac{J^T \Lambda^T D^{-1}\Lambda J}{4}\\ &= \; det(\Lambda) \sqrt{\frac{(\pi)^N}{ det(K)}} e^\frac{J^T K^{-1} J}{4}\quad \quad \quad = \; det(\Lambda) \sqrt{\frac{(\pi)^N}{ det(K)}} e^\frac{\sum_{ij}J_i K^{-1}_{ij} J_j}{4}.
    \end{align*}
 
 \bigskip
    
Taking the logarithm of $Z[J]$
\begin{align*}
    \ln(Z[J])=    \ln\left( det(\Lambda) \sqrt{\frac{(\pi)^N}{ det(K)}}\right)  + \frac{\sum_{ij}J_i K^{-1}_{ij} J_j}{4}
\end{align*}

Finally leading to:

\begin{equation}
\boxed{
    \frac{\partial^2\ln(Z)}{\partial J_k \partial J_l}= \frac{1}{2} K^{-1}_{kl}
    }
\end{equation}

\clearpage

\section{Pseudo-inverse of the Laplacian for the networks.}
\label{appendix:b}
In this appendix, we present the density plots of the pseudo-inverse of the Laplacian for each network.

\begin{center}\
\begin{figure}[H]\
\centering
    \includegraphics[width=0.95\textwidth]{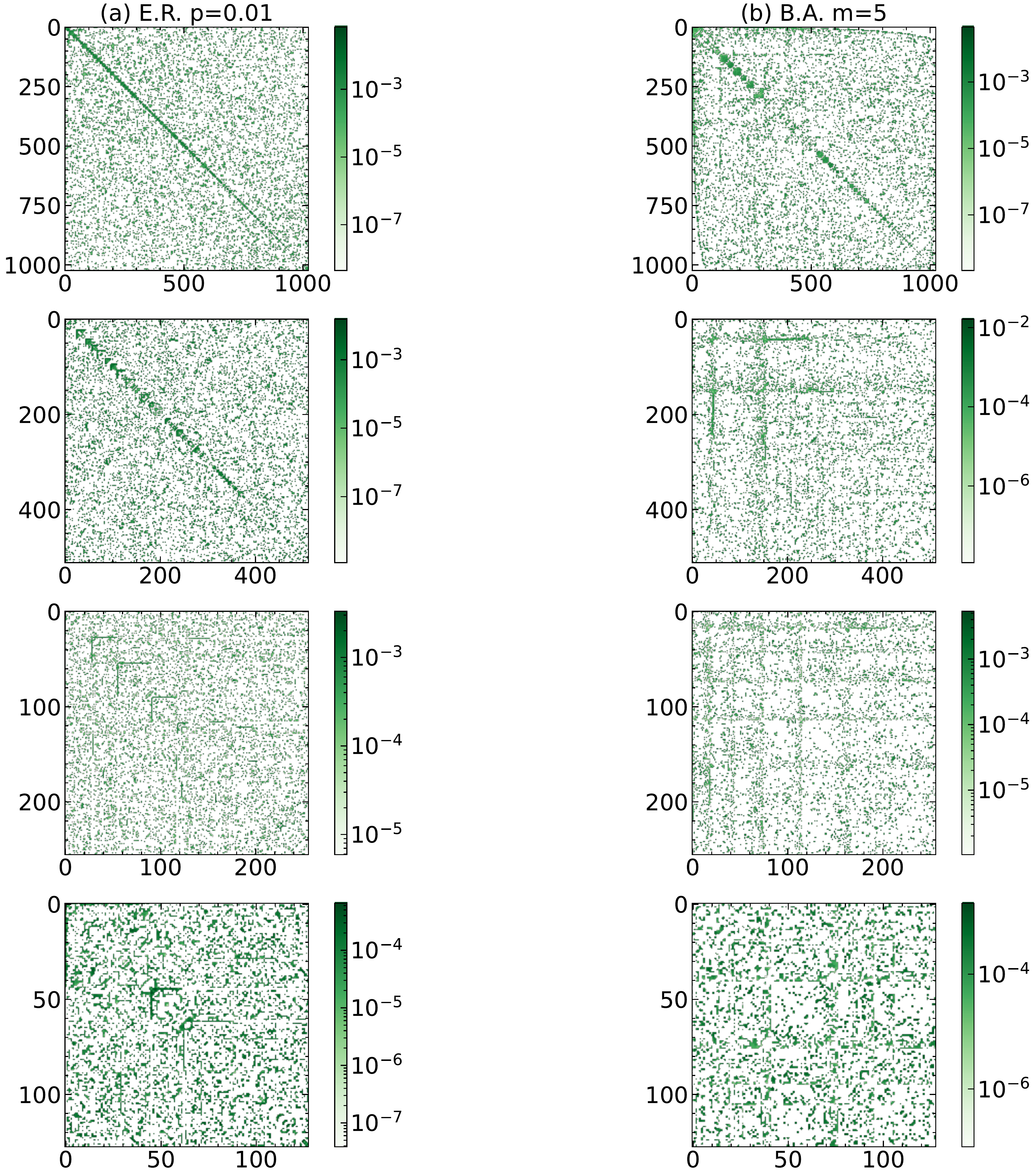}\hfill

    \caption{(Color online.) Density plot of the pseudo-inverse of the Laplacian in each artificial network for the positive correlations.}
    \label{fig:pinv_art}
\end{figure}\
\end{center}\

\begin{center}\
\begin{figure}[H]\
\centering
    \includegraphics[width=1.0\textwidth]{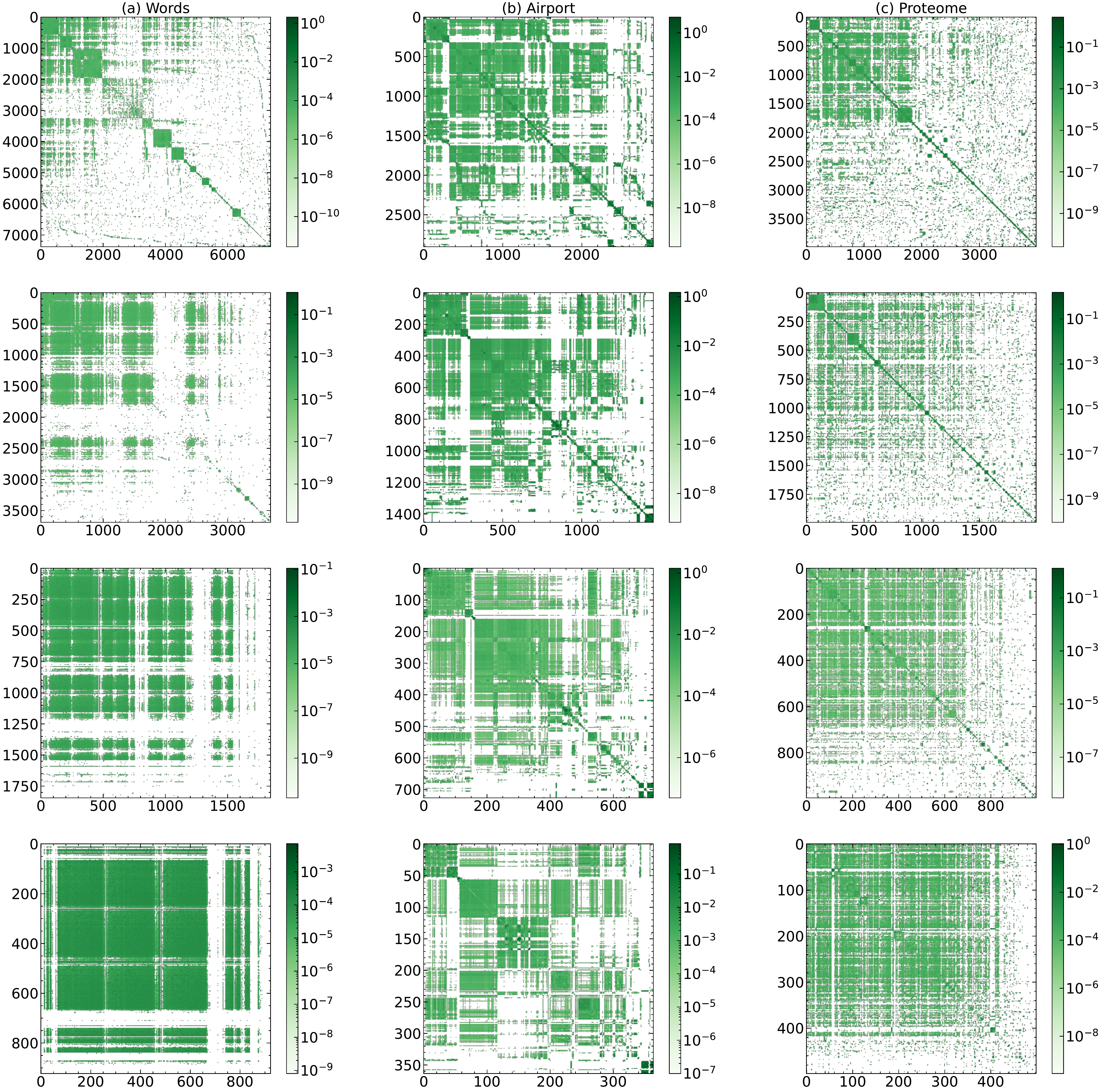}\hfill

    \caption{(Color online.) Density plots of pseudo-inverse of the Laplacian  in Words, Airport, and  Proteome networks for the positive correlations. }
    \label{fig:pinv_real1}
\end{figure}\
\end{center}\

\begin{center}
\begin{figure}[H]
\centering
    \includegraphics[width=1.0\textwidth]{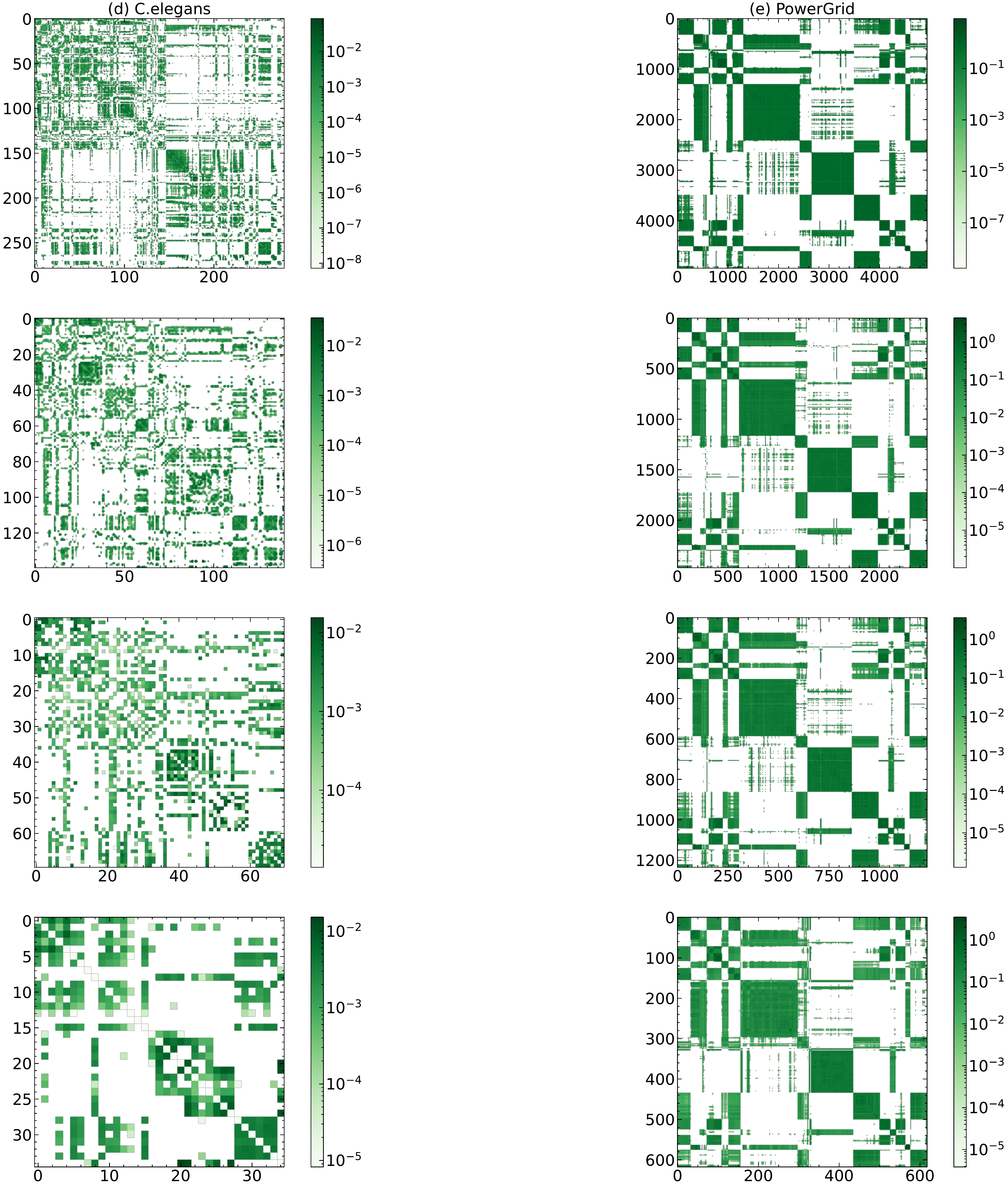}\hfill

    \caption{(Color online.) Density plots of pseudo-inverse of the Laplacian  in  \textit{C.elegans} and Power Grid networks for the positive correlations.}
    \label{fig:pinv_real2}
\end{figure}
\end{center}

\newpage

\nocite{*}
\end{appendices}

\bibliography{bibliography}

\end{document}